\documentclass[aps,prl,twocolumn,groupedaddress,showpacs]{revtex4}

\usepackage{amsmath}    
\usepackage{amsfonts}
\usepackage{amssymb}
\usepackage{graphicx}   

\newcommand{\R}{$\mathbb{R}$}
\newcommand{\UD}{1D}
\newcommand{\DD}{2D}
\newcommand{\TD}{3D}
\newcommand{\FD}{4D}

\newcommand{\WTMM}{WTMM}
\newcommand{\WTMMM}{WTMMM}
\newcommand{\WT}{WT}
\newcommand{\gra}[1]{{\mathbf #1}}
\newcommand{\ron}[1]{{\mathcal #1}}
\newcommand{\bpsi}{\boldsymbol{\psi}}
\newcommand{\gb}{\gra{b}}
\newcommand{\Mpsi}{\ron{M}_{\bpsi}}
\newcommand{\Tpsi}{{\mathbf T}_{\bpsi}}

\newcommand{\carreBlanc}{\protect\scalebox{0.6}{\ensuremath{\square}}}
\newcommand{\rondBlanc}{\ensuremath{\circ}}
\newcommand{\carreNoir}{\protect\scalebox{0.6}{\ensuremath{\blacksquare}}}

\newcommand{\conv}{\ensuremath{*}}

\newcommand{\triangleNoir}{\protect\scalebox{0.9}{\ensuremath{\blacktriangle}}}

\begin{document}


\title{A three-dimensional wavelet based multifractal method : about the need
of revisiting the multifractal description of turbulence dissipation data}


\author{Pierre Kestener}
\author{Alain Arneodo}
\affiliation{Laboratoire de Physique, Ecole Normale Sup\'erieure de Lyon, 46 all\'ee d'Italie, 69364 Lyon c\'edex 07, France}

\date{October 24, 2003}

\begin{abstract}
We generalize the wavelet transform modulus
 maxima (\WTMM{}) method to multifractal analysis of \TD{} random fields.
This method is calibrated
on synthetic \TD{} monofractal fractional Brownian fields and on 
\TD{} multifractal singular cascade measures as well as 
their random function counterpart obtained by fractional integration. 
Then we apply
the \TD{} \WTMM{} method to the dissipation 
field issued 
from \TD{} isotropic turbulence simulations. 
We comment on the need to revisiting previous box-counting analysis which have
 failed to estimate correctly the corresponding multifractal spectra because
of their intrinsic inability to master non-conservative singular cascade measures.
\end{abstract}


\maketitle

The multifractal formalism was introduced in the mid-eighties to provide a
 statistical description of the fluctuations of regularity of singular
measures found in chaotic dynamical
 systems~\cite{ref1} or in modelling 
the energy cascading process in turbulent 
flows~\cite{ref2,aMen91,bFri95}.
Box-counting (BC) 
algorithms were successfully adapted to resolve
multifractal scaling for isotropic self-similar 
fractals~\cite{aMen91,ref5}.
As to self-affine fractals, Parisi and Frisch~\cite{aPar85} developed, 
in the context of turbulence velocity data analysis, an
alternative multifractal description based on 
the so-called structure functions.
Unfortunately,  there are
some drawbacks to these classical multifractal methods and as proposed in
Ref. \cite{ref9},
%
a natural way of performing a unified multifractal analysis of both singular 
measures and multi-affine functions, consists in using the 
\textit{continuous wavelet transform} (\WT{}). 
Applications of the so-called \WTMM{} method to \UD{} signals have already
 provided insight into 
a wide variety of problems, \textit{e.g.} fully-developed turbulence, 
financial markets, meteorology, physiology, DNA sequences~\cite{bArn95}.
Recently, the \WTMM{}
method has been generalized to \DD{} for multifractal analysis
of rough surfaces~\cite{ref14},
and successfully applied to characterize
the intermittent nature of satellite images of the cloud 
structure~\cite{ref15} and to assist in the diagnosis in
digitized mammograms~\cite{akes01}.
Our aim here is to go one step further and to generalize the \WTMM{} method 
from \DD{} to \TD{}.

The \TD{} \WTMM{} method consists in 
smoothing the 
discrete \TD{} field data by convolving it with a filter and then in computing
the gradient on the smoothed signal as for
multiscale edge detectors~\cite{ref10}. Define three wavelets 
$\psi_i(x,y,z)=\partial \phi(x,y,z)/\partial x_i$ with $x_i = x, y$ or $z$
 for $i = 1,2$ or 3 respectively and $\phi (x,y,z)$ is a \TD{} smoothing
 function well localized around $x=y=z=0$. For any function 
$f(x,y,z)\in L^2$(\R$^3$), the \WT{} 
at the point $\gra{b}$
 and scale $a$
can be expressed in a vectorial form~\cite{ref14,ref10}:
\begin{equation}
  \Tpsi [f] (\gb,a)
    = {\boldsymbol \nabla} \{ \phi_{{\mathbf b},a} * f \} \;,
\label{eq1}
\end{equation}
where $\phi_{{\mathbf b},a}(\gra{r})=a^{-3}\phi(a^{-1}(\gra{r}-\gra{b}))$. If
 $\phi$ is just a Gaussian $\phi(\gra{r})=\exp (-\gra{r}^2/2)$ (or
one of its derivatives), then Eq.~(\ref{eq1}) defines the \TD{} \WT{} as the gradient field vector of 
$f(\gra{r})$ smoothed by dilated versions $\phi(\gra{r}/a)$ of this 
filter.
At a given scale $a$, the \WTMM{} are defined by the positions $\gra{b}$ where
the \WT{} modulus 
$\Mpsi [f](\gb,a)= | \gra{T}_{\bpsi}[f]({\mathbf b},a) |$ 
is locally maximum along the 
 direction 
 of the \WT{} vector (Eq. (\ref{eq1})).
These \WTMM{} lie on connected surfaces  called 
\textit{maxima surfaces} (see Fig. \ref{fig2}).
In theory, at each scale $a$, one only needs
 to record the position of the local maxima of $\Mpsi$ (\WTMMM)
along the maxima surfaces together with the value of $\Mpsi[f]$ and the \WT{}
vector direction.
They indicate locally the direction where the signal
has the sharpest variation. 
 These \WTMMM{} are disposed along connected curves across scales called {\em
  maxima lines}~\cite{ref10,ref14} 
living in a \FD{} space
$(x,y,z,a)$.  We will define the \WT{}
skeleton as the set of maxima lines that converge to the $(x,y,z)$-hyperplane
in the limit $a \rightarrow 0^+$ (Fig \ref{fig2}(d)).
One can prove~\cite{ref10,ref14} that,
 provided the first $n_{\bpsi}$ moments of $\bpsi$ be zero, then
$\Mpsi[f] \bigl(\ron{L}_{{\mathbf r_0}}(a) \bigr)\sim 
    a^{h({\mathbf r_0})}$
along the maxima line $\ron{L}_{{\mathbf r_0}}(a)$ pointing to the point 
$\gra{r}_0$ in the limit $a \rightarrow 0^+$, where $h(\gra{r}_0)$
($<n_{\bpsi}$)  
is the local H\"older exponent of $f$.
As in \UD{} and \DD{} \cite{ref9,ref14}, the \TD{} \WTMM{} method
consists in defining the partition functions :
\begin{equation}
{\mathcal Z}(q,a)=\sum_{{\mathcal L}\in {\mathcal L}(a)} \left ( 
\ron{M}_{\bpsi}[f]({\mathbf r},a)\right)^q  \sim a^{\tau(q)},
\label{eq3}
\end{equation}
where $q \in$ \R\; and ${\mathcal L}(a)$ is the set of maxima 
lines of the \WT{} skeleton.
%
Then from Legendre transforming 
 $\tau(q)$: $D(h)=\min_q \bigl( qh-\tau(q) \bigr)$,
one gets the $D(h)$ singularity spectrum defined as the Hausdorff dimension of the
set of points $\gra{r}$ where $h(\gra{r})$ is $h$.
As an alternative strategy one can compute
the mean quantities $h(q,a)$ and $D(q,a)$~:
\begin{align}
h(q,a)=&
\sum_{{\mathcal L}\in{\mathcal L}(a)} \ln \left| 
\ron{M}_{\bpsi}[f]({\mathbf r},a) \right| \; W_{\bpsi}[f](q,{\mathcal L}, a)\; ,
\label{eq5}\\
D(q,a)=&
\sum_{{\mathcal L}\in{\mathcal L}(a)} W_{\bpsi}[f](q,{\mathcal L}, a) \; \ln
\bigl( W_{\bpsi}[f](q,{\mathcal L}, a) \bigr) \; ,
\label{eq6}
\end{align}
where 
$W_{\bpsi}[f](q,{\mathcal L}, a)=\bigl( \ron{M}_{\bpsi}[f]({\mathbf r},a) \bigr)^q/{\mathcal Z}(q,a)$
is a Boltzmann weight computed from the \WT{} skeleton. From the scaling
behavior of these quantities, one can extract $h(q) =\lim_{a\rightarrow 0^+} h(q,a)/\ln a$ and $D(q) =\lim_{a\rightarrow 0^+} D(q,a)/\ln a$
and therefore the $D(h)$ spectrum.

\begin{figure}
  \centering
  \scalebox{0.48}{\includegraphics[]{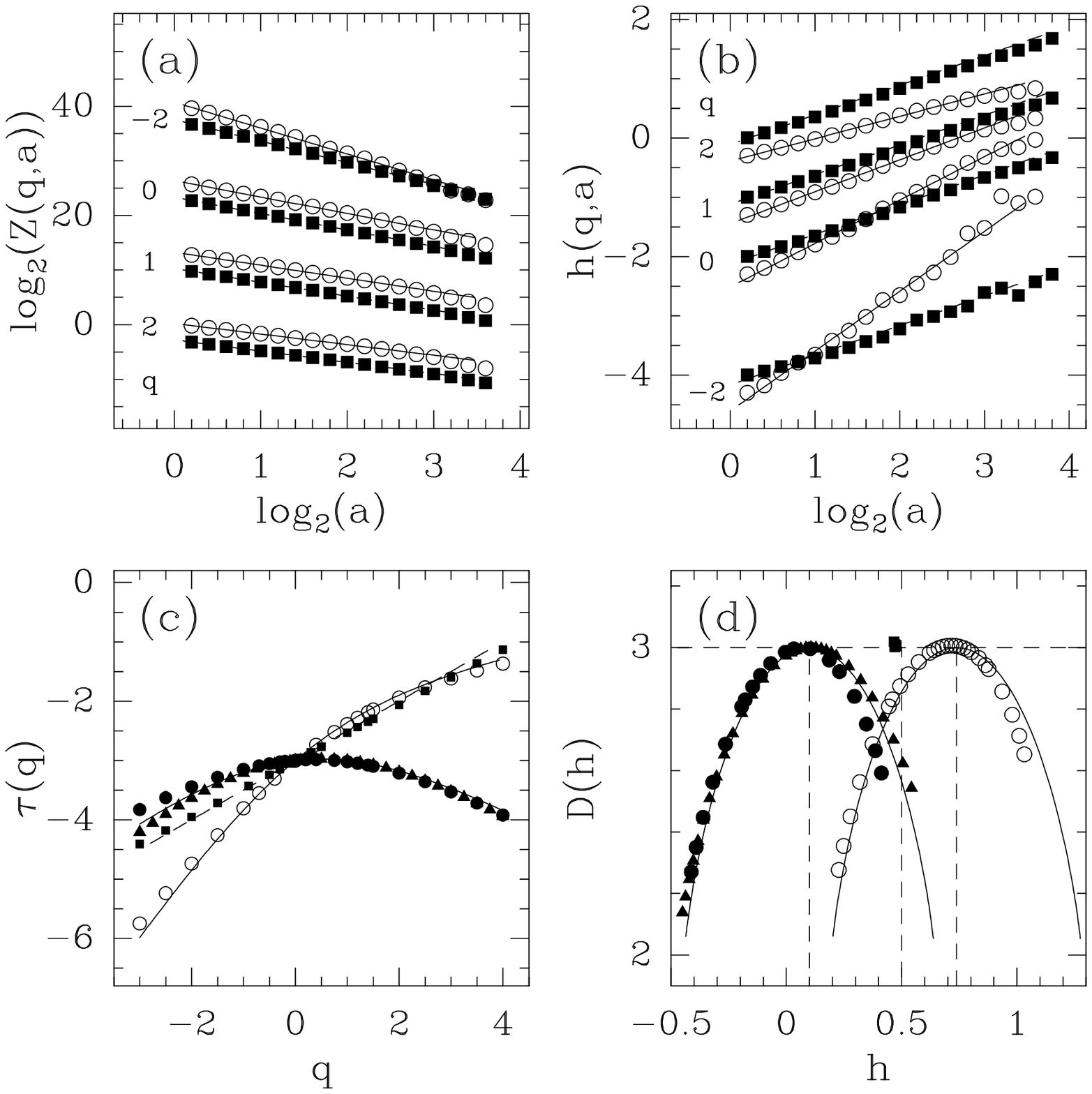}}
  \caption{
    Multifractal analysis of \TD{} $B_{H=1/2}$ ($\carreNoir$), \TD{} p-model
    ($\bullet$, $p=0.32$) and \TD{} FISC ($\circ$, $p=0.32$, $H^*=0.638$)
    data using the \TD{} \WTMM{} method with a first ($\carreNoir$) or
    third ($\bullet$,$\circ$) order analyzing wavelet ($\phi(\gra{r})$ is
    a Gaussian or its Laplacian).
    (a) $\log_2 {\cal Z}(q,a)$ vs $\log_2 a$; 
    (b) $h(q,a)$ vs $\log_2 a$;
    the dashed and solid lines correspond 
    to linear regression fits over the three first octaves.
    (c) $\tau(q)$ vs $q$; the dashed and solid lines
     are the theoretical predictions.
    (d) $D(h)$ vs $h$ as obtained from the
    scaling behavior of $h(q,a)$ and $D(q,a)$.
    These results correspond to annealed averaging over $16$ $\times$
    $(256)^3$ images. $a$ is expressed in $\sigma_W$ units where $\sigma_W
    =7$ (pixels) is the characteristic size of $\bpsi$ at the smallest
    resolved scale.
    In (c) and (d), the symbols ($\triangleNoir$) represent the
    results of classical box-counting analysis of the 3D p-model.  
    \label{fig1}
  }
\end{figure}

\begin{figure}
  \begin{minipage}[c]{.45\linewidth}
    \scalebox{0.2}{
      \includegraphics{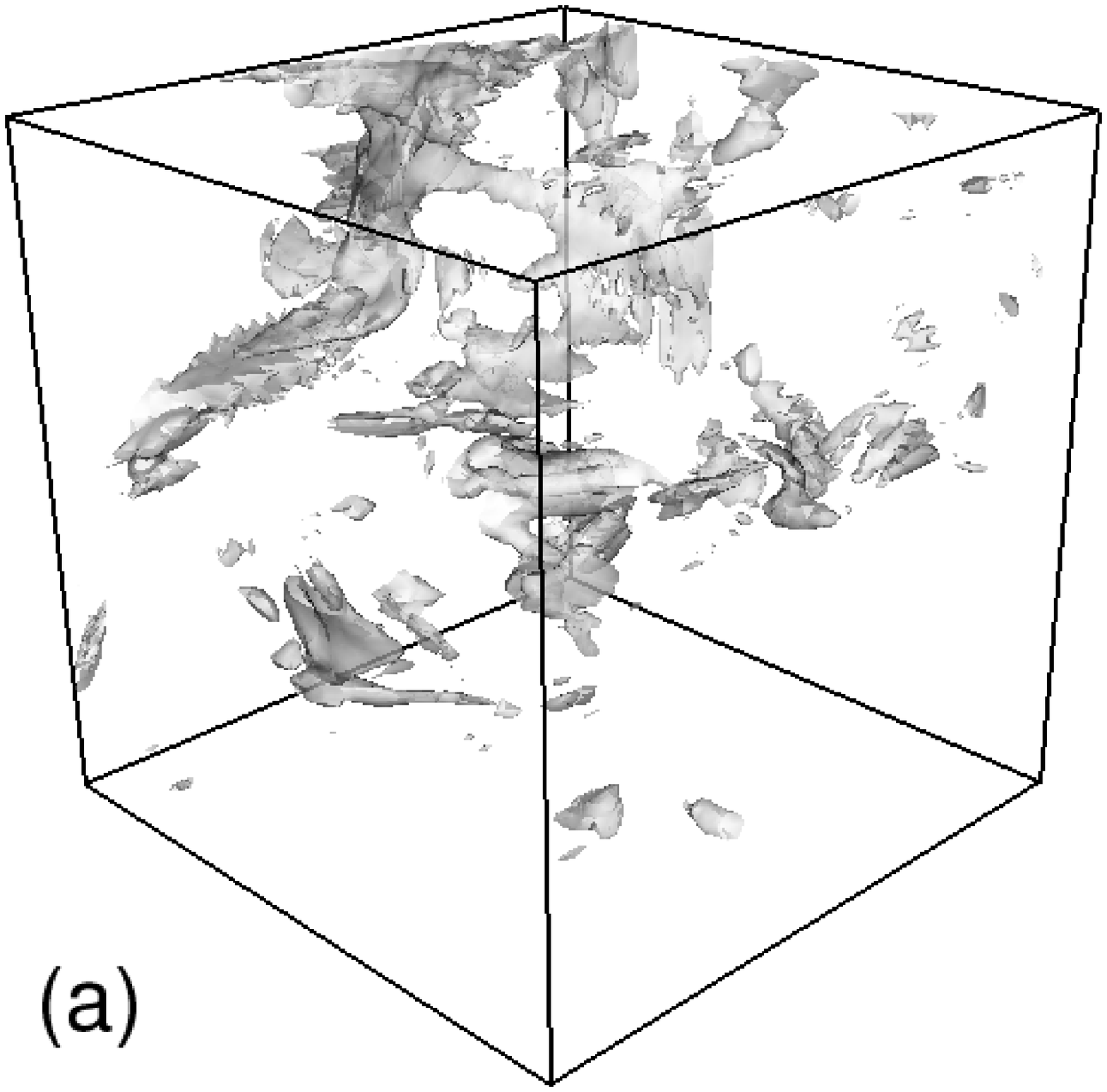}
      }
  \end{minipage}
  \hfill
  \begin{minipage}[c]{.45\linewidth}
    \scalebox{0.24}{
      \includegraphics{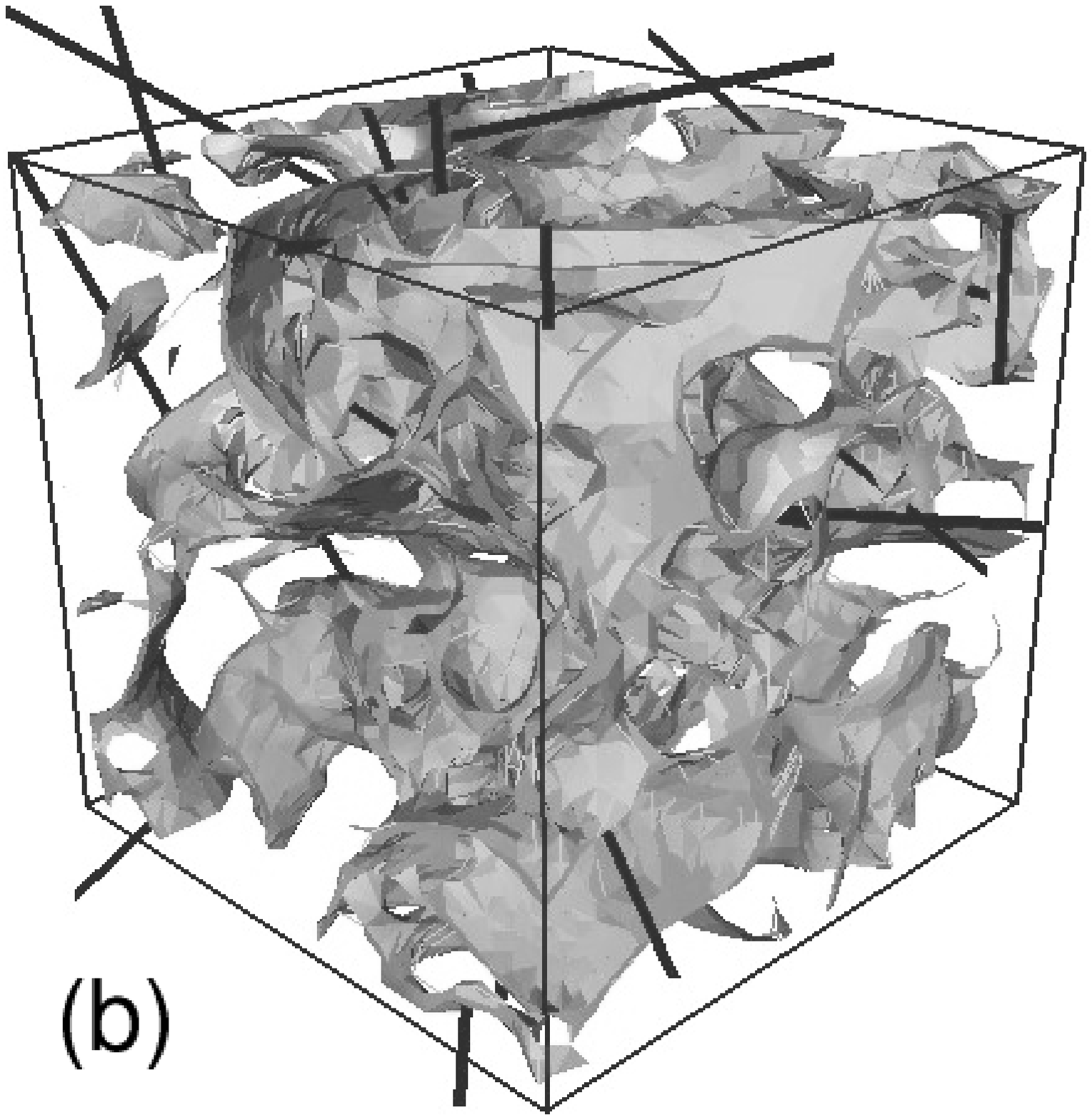}
      }
  \end{minipage}
  \begin{minipage}[c]{.45\linewidth}
    \scalebox{0.24}{
      \includegraphics{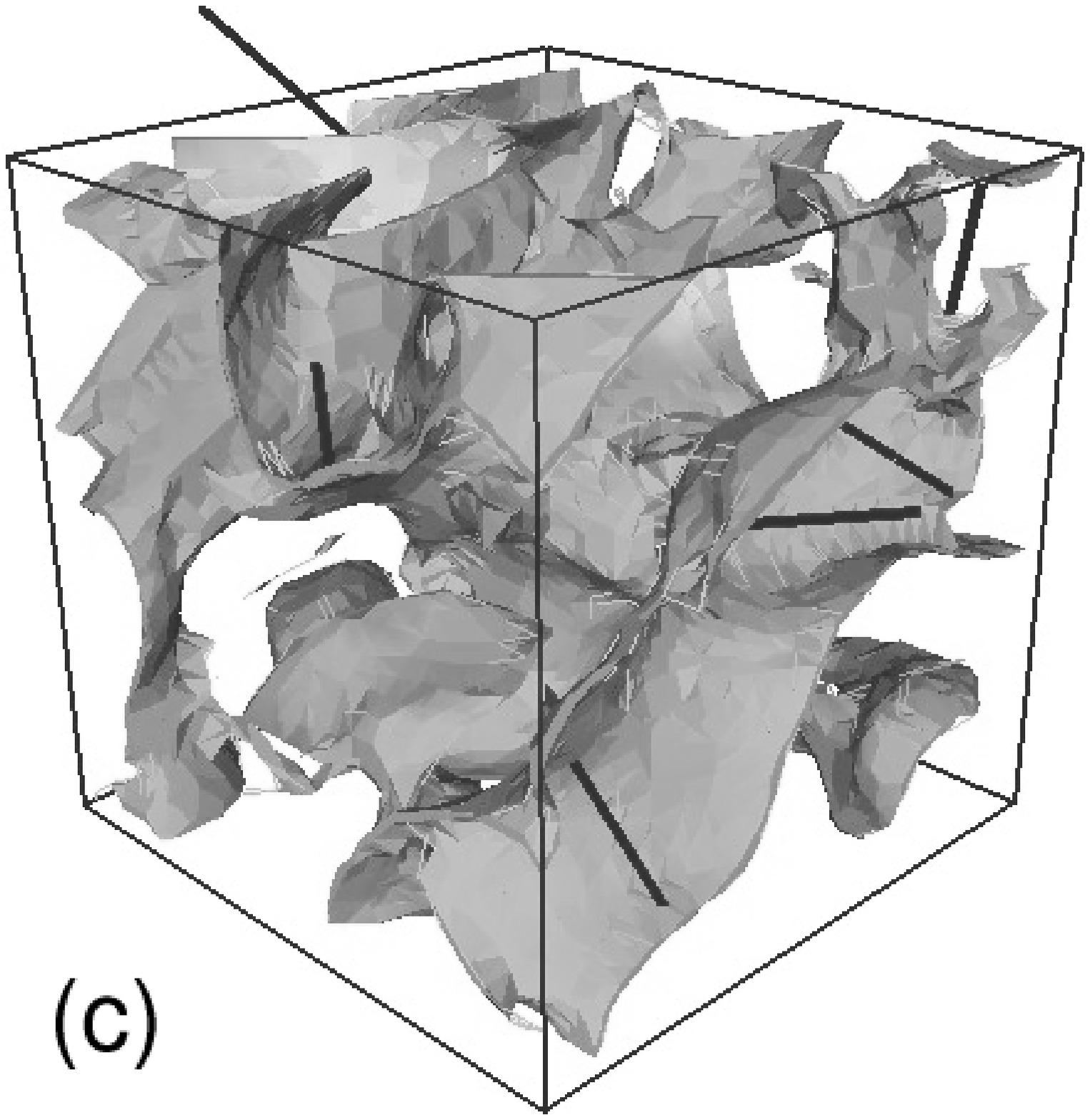}
      }
  \end{minipage}
  \hfill
  \begin{minipage}[c]{.45\linewidth}
    \scalebox{0.22}{
      \includegraphics{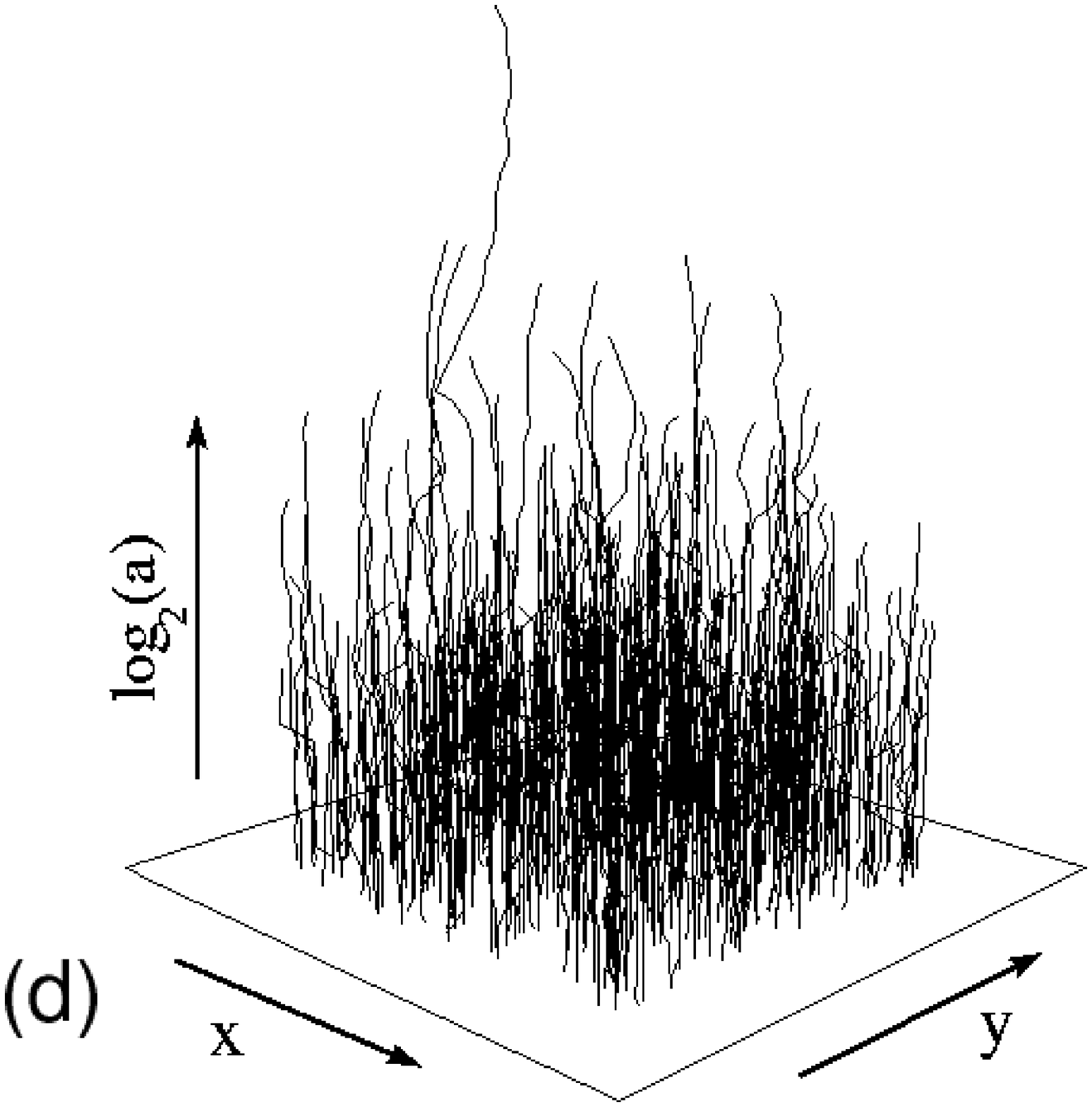}
      }
  \end{minipage}
  \caption{
        \TD{} \WT{} analysis of the DNS dissipation field $\epsilon$.
         $\bpsi$ is the first-order
        analyzing wavelet ($\phi(\gra{r})$ is the Gaussian).
        (a) 
        Isosurface plot of $\epsilon$ in a $(64)^3$ 
        sub-cube.
        (b) \WTMM{} surfaces at scale $a=2\sigma_W$; from the
         local maxima (\WTMMM{}) of $\Mpsi$ along these surfaces originate a
        black segment whose length is proportional to $\Mpsi$ and direction
        is along the \WT{} vector.
        (c) same as in (b) for $a=4\sigma_W$.
        (d) \TD{} projection of the \WT{} skeleton obtained by 
        linking the \WTMMM{} across scales.
    \label{fig2}
  }
\end{figure}

Fractional Brownian motions (fBm) are homogenous random self-affine functions
 that have been 
used to calibrate both the \UD{} and \DD{} \WTMM{}
methods~\cite{ref9,ref14}.
\TD{} fBm $B_H(\gra{r})$ are Gaussian stochastic processes with stationary 
increments and well-known statistical properties : 
$\tau(q)=qH-3 \; , \; 0<H<1 $.
In Fig. 1 are reported the results of the \TD{} \WTMM{} method when applied 
to 16 $\times$ $(256)^3$ realizations of $B_{H=1/2}(\gra{r})$. As shown in Fig. 1(a),
${\mathcal Z}(q,a)$ (Eq. (\ref{eq3})) display nice
 scaling behavior over 3 octaves
when plotted \textit{vs} $a$ in a logarithmic representation.
A linear 
regression fit of the data
 yields
 the linear $\tau(q)$ spectrum shown in Fig. 
\ref{fig2}(c),
in good agreement with the theoretical spectrum.
This signature of monofractality is confirmed in Fig. \ref{fig1}(b) where, 
when plotting $h(q,a)$ vs $\log_2(a)$,
$H=1/2$ is shown to provide an excellent fit of the slope of the data
($\sigma_W \lesssim a \lesssim 4\sigma_W$)
for
 $q \in ]-2,4[$.
When using $h(q,a)$ (Eq. (\ref{eq5})) and
$D(q,a)$ (Eq. (\ref{eq6})) to estimate $h(q)$ and $D(q)$ in
Fig. \ref{fig1}(d), 
one
gets, up to the numerical uncertainty, a $D(h)$ spectrum that
reduces to a single
 point $D(h=H=1/2)=3$.
Similar quantitative estimates 
have been obtained for 
$H=1/3$ and $2/3$,
thus confirming that \TD{} fBm's are nowhere differentiable with
a unique H\"older exponent $h=H$.

Generating multifractal measures using multiplicative cascades is  well
documented~\cite{ref2,aMen91,bFri95}.
The ``binomial (or p-) model'',
originally designed
to account for  the statistical scaling properties of the dissipation field
 in fully developed turbulence, has very simple multifractal 
properties~\cite{ref2,aMen91,bFri95}.
The \TD{} version of the p-model consists in starting with 
a cube of 
size $L$, in which a measure $\mu=\mu_L$ is uniformly distributed.
At the first step, the initial cube is broken into eight
smaller cubes of size $L/2$ and
one selects
at random 
the four sub-cubes which will receive a fraction
$M^{(1)}=p_1/4$, the four others receiving the fraction $M^{(2)}=p_2/4$ of the
measure 
 ($p_1+p_2=1$).
Iterating this rule,
one generates a random
 singular measure 
 $\mu_n(\gra{r};l) = \mu_L \prod_{i=1}^n M_i \; , \; l/L=2^{-n} \rightarrow 0$.
%
%
%
%
A straightforward computation 
yields : $\tau_\mu(q)=-(q+2)-\log_2 \bigl( p_1^q+p_2^q \bigr)$.
As a mean of introducing continuity, Fractionaly Integrated Singular Cascade
(FISC) algorithm~\cite{aSch87} amounts to a low-pass power-law
 filtering in Fourier space:
$f_n(\gra{r})=\mu_n(\gra{r}) \conv |\gra{r}|^{-(1-H^*)}\; , \; 0< H^* < 1$.
This leads to 
FISC 
random functions with the following multifractal spectrum :
 $\tau_f(q) =\tau_\mu(q)+qH^* = -2-q(1-H^*)-\log_2 \bigl( p_1^q+p_2^q \bigr)$.
In Fig. \ref{fig1} are reported the results of the \TD{} \WTMM{} analysis of 
the FISC model with 
$p_1=0.32$ and $H^*=0.638$.
As shown in Fig. \ref{fig1}(a), 
${\mathcal Z}(q,a)$ 
display good scaling for $q\in ]-2, 4[$ for which statistical convergence turns
out to be achieved.
The corresponding $\tau_{f}(q)$ spectrum is displayed in Fig. \ref{fig1}(c)
along with the theoretical spectrum. The agreement is 
quite satisfactory; $\tau_{f}(q)$ is nonlinear,
 the hallmark of multifractal scaling.
This is confirmed in Fig. \ref{fig1}(b) where the slope of $h(q,a)$ 
vs $\log_2(a)$ clearly depends on $q$. From the estimate of 
$h(q)$ and
$D(q)$,
one gets the
single-humped $D_f(h)$ spectrum shown in Fig. \ref{fig1}(d).
 Note that some departure from the theoretical spectrum 
can be observed on the right-hand side
of the $D(h)$-curve ($q\lesssim -2$) indicating that  
the estimate of the weakest 
singularities would require a larger statistical sample.
In Figs. \ref{fig1}(c) and \ref{fig1}(d) are also reported the results of
a comparative analysis 
of the p-model using the 3D WTMM method and classical
BC techniques.
Note that the \WTMM{}
definition of the $\tau_{\mu}^{WT}(q)$ spectrum (Eq. (\ref{eq3})) slightly
differs from the BC definition $\tau_{\mu}^{BC}(q)$ found
in the literature~\cite{aMen91} : 
\begin{equation}
\tau_{\mu}^{BC}(q) = \tau_{\mu}^{WT}(q) +dq = (q-1)D_q \;,
\label{eq11}
\end{equation}
which implies the following relationship 
$f_{\mu}^{BC}(\alpha) = D_{\mu}^{WT}(h=\alpha-d)$
 between the corresponding singularity spectra ($d=3$).
For both methods, the numerical results for
$\tau_\mu(q)$ and $D_\mu(h)$ are found in good agreement with the
 theoretical spectra,
 for $q \in ]-2, 4[$. In particular, the cancellation
exponent~\cite{ref26} is found
$\tau_\mu^{BC}(q=1)=\tau_\mu^{WT}(q=1)+3=0$, as the  
signature of the conservativity of the p-model cascading 
rule. One of the main problem with the BC method is
the fact that, by construction, the measure in a given box is the sum of
the measures in smaller non-overlapping boxes, which implies that the
cancellation exponent $\tau_\mu^{BC}(q=1)=0$.
This means that BC algorithms are not adapted to study
non-conservative singular cascades, signed measures as well as
multifractal functions for which the cancellation exponent
has no reason to vanish.
Altogether the results reported in Fig.~1 bring the
demonstration that our \TD{} 
\WTMM{} methodology paves the way from multifractal analysis of singular
measures to continuous multi-affine functions.
In particular the $D_f(h)$ curve for the FISC random function is found
identical  to the $D_\mu(h)$ 
curve up to a translation to the right by $H^*$; this is the consequence
of the fractional integration which implies $D_f(h)=D_\mu(h-H^*)$.

Since Kolmogorov's founding work (K41), fully developed
 turbulence has been intensively studied theoretically, numerically and 
experimentally~\cite{aMen91,bFri95}. A central quantity in the K41
theory is the mean energy dissipation $\epsilon = \frac{\nu}{2} \sum_{i,j} (\partial_j v_i + \partial_i v_j)^2$ 
 which is supposed to be 
constant.
Indeed, $\epsilon$ is not spatially homogenous but
undergoes local intermittent fluctuations~\cite{aMen91,bFri95}.
There have been early experimental attemps to measure the multifractal spectra
of $\epsilon$~\cite{aMen91}.
Surprisingly, 
the binomial model turns out 
to account reasonably well
 for the observed $\tau_{\epsilon}(q)$ and $f_{\epsilon}(\alpha)$ 
spectra~\cite{aMen91}. 
Experimentally,
 single probe measurement of the longitudinal
velocity requires the use of the \UD{} surrogate dissipation approximation
$\epsilon'=15\nu (\partial u / \partial x)^2$ that may
 introduce severe bias in the
multifractal analysis.
\TD{} multifractal processing of dissipation data is at
 the moment feasible only for numerically simulated flows 
at moderate Reynolds number 
for which scaling just begins to manifest itself~\cite{ref20}.
Several numerical studies~[16a,c] agree that
$\epsilon'$ is in general more intermittent
 than $\epsilon$
 which is found nearly 
log-normal in the inertial range~[16b,c].

\begin{figure}[b]
  \centering
  \scalebox{0.48}{\includegraphics[]{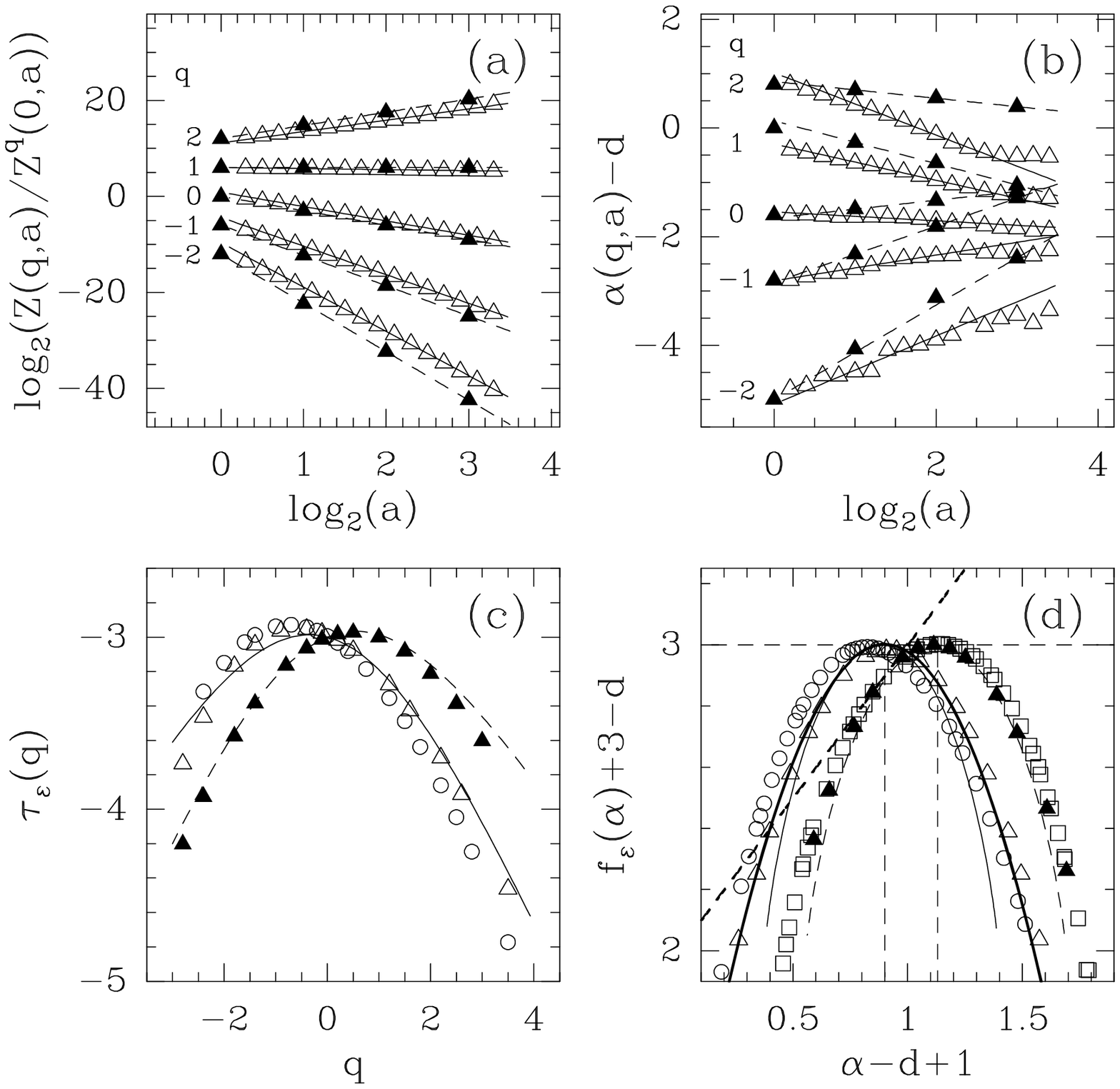}}
  \caption{
        Multifractal analysis of Meneguzzi DNS dissipation 
        data ($d=3$)
        using the \TD{} \WTMM{} method 
        ($\triangle$) and BC techniques ($\blacktriangle$).
        (a) $\log_2 ({\cal Z}(q,a)/({\cal Z}(q=0,a))^q)$ vs $\log_2 a$; 
        (b) $h(q,a)=\alpha(q,a)-3$ vs $\log_2 a$;
        the solid and dashed lines correspond 
        to linear regression fits over
        $\sigma_W \lesssim a \lesssim 2^3\sigma_W$.
        (c) $\tau_{\epsilon}(q) = \tau_{\epsilon}^{WT}(q)$ 
         or $\tau_{\epsilon}^{BC}(q)-3q$
        vs $q$; the solid (dashed) lines
        correspond to the p-model prediction with $p_1=0.36$, $p_2=0.78$
        ($p_1=0.32$, $p_2=0.68$).
        (d) $f_{\epsilon}(\alpha)+3-d$ vs $\alpha-d+1$, where 
        $f_{\epsilon}(\alpha) = f_{\epsilon}^{BC}(\alpha)$ 
        or $f_{\epsilon}^{WT}(\alpha) = D_{\epsilon}^{WT}(h=\alpha-d)$ 
        ; the solid and dashed lines have the same meaning as in
        (c); the thick solid line is the log-normal
        spectrum $f_{\epsilon}(\alpha)=d-(\alpha-d+1-C_1)^2/2C_2$ with
        $C_1=0.91$ and $C_2=0.22$. The symbols ($\carreBlanc$) correspond
        to some average $f(\alpha)$ spectrum of experimental ($d=1$)
        surrogate dissipation data~\cite{aMen91}.
        In (c) and (d), the 3D WTMM multifractal spectra of
        L\'ev\^eque 3D dissipation data (18 snapshots) ($\rondBlanc$)
        are shown for comparison.
        In (d) the dashed straight line is the diagonal. 
        \label{fig3}
      }
\end{figure}

So far mainly BC 
techniques have been used to perform multifractal
analysis of numerical and experimental dissipation data~\cite{aMen91,ref20}.
Here we apply the \TD{} \WTMM{} method to 
isotropic turbulence DNS data obtained by Meneguzzi with the same 
numerical code as 
in Ref. \cite{aMene} but at a
$(512)^3$ resolution
and a viscosity of $5.10^{-4}$
corresponding to a Taylor
Reynolds number $R_\lambda=216$ (one snapshot of the dissipation
\TD{} spatial field).
%
For the sake of comparison, we will also report the results of some
averaging over 18 snapshots of ($256$)$^3$ DNS run by L\'ev\^eque at
$R_\lambda = 140$.
The main steps of our \TD{} \WT{} computation 
 are
illustrated in Fig. \ref{fig2}. Focusing on a ($64^3$) sub-cube, we show the
original $\epsilon$ data 
(Fig. \ref{fig2}(a)), 
the \WTMM{} surfaces along with the \WTMMM{} points computed
at two different scales (Figs. \ref{fig2}(b) and \ref{fig2}(c)) and some 
projection of the \WT{} skeleton (Fig. \ref{fig2}(d)).
According to Eq. (\ref{eq11}), by plotting ${\mathcal Z}(q,a) / ({\mathcal
  Z}(q=0,a))^q$ \textit{vs} $a$  
in a logarithmic
representation in Fig. \ref{fig3}(a), one can then directly compare
(${\mathcal Z}(q=0,a) \sim a^{-d}$) our \WTMM{} computations with
BC ones.
Actually, good scaling properties are observed for $q \in ]-2, 4[$.
Linear regression fits of the data yield the non-linear 
$\tau_{\epsilon}(q)$ spectra shown in Fig. \ref{fig3}(c) that significantly
deviate from a straight line, the hallmark of multifractality.
But surprisingly,
$\tau_{\epsilon}^{WT}(q)$
significantly differs from 
$\tau_{\epsilon}^{BC}(q)-3q$.
Actually, 
our \TD{} \WTMM{} algorithms reveal that the cancellation exponent~\cite{ref26}
 is significantly different from zero : 
$\tau_{\epsilon}^{WT}(q=1) + 3 = -0.19 \pm 0.03 <0$,
the signature of a signed measure. 
Indeed, as shown in Fig. \ref{fig3}(c), 
$\tau_{\epsilon}^{WT}(q)$ data are rather nicely fitted
by the theoretical spectrum $\tau_\mu(q)$ 
 of the non-conservative p-model with $p_1=0.36$ and
$p_2=0.78$ ($p_1+p_2=1.14 >1$). 
Thus BC 
algorithms systematically provide a misleading
 conservative $\tau_{\epsilon}(q)$ spectrum diagnostic 
with $p=p_1/(p_1+p_2)$ and
$1-p=p_2/(p_1+p_2)$.
As shown in Fig. \ref{fig3}(c), $\tau_{\epsilon}^{BC}(q)-3q$
 data are quite well
reproduced by the theoretical conservative p-model spectrum with
$p=0.32=p_1/(p_1+p_2)=0.36/1.14$, consistently with our \TD{} \WTMM{}
finding for $\tau_{\epsilon}^{WT}(q)$.
The
difference between the two spectra is nothing but a fractional integration 
of exponent $H^* = \log_2(p_1+p_2) \sim 0.19$.  This result is confirmed in
Fig. \ref{fig3}(d) where the singularity spectrum
$f_{\epsilon}^{BC}(\alpha)$  
is misleading shifted to the right
by $H^*$ ($= -$ the cancellation exponent) $\sim 0.19$, without
any shape change as compared to the  
$f_{\epsilon}^{WT}(\alpha) = D_{\epsilon}^{WT}(h=\alpha-d)$.
Note that the observation that $f_\epsilon^{WT}(\alpha-d+1)$ is not
tangent to the diagonal (on the contrary to $f_\epsilon^{BC}(\alpha-d+1)$)
is clearly related to the fact that the cancellation exponent is
different from zero~\cite{aZhou01}.
This observation seriously questions the validity of most of the experimental
and numerical BC estimates of the $\tau_{\epsilon}^{BC}(q)$ and
$f_{\epsilon}^{BC}(\alpha)$ 
spectra reported so far in the
 literature. In Fig. \ref{fig3}(d) is shown for comparison some average 
$f_{\epsilon}^{BC}(\alpha)$ spectrum obtained by Meneveau and
 Sreenivasan~\cite{aMen91} for 
\UD{} surrogate dissipation 
experimental data ($d=1$).
 We notice that these experimental data were claimed to be
well fitted by the conservative p-model with $p=0.3$ ($1-p=0.7$), \textit{i.e.}
a value not so far from the value $p=p_1/(p_1+p_2)=0.32$ derived from our 
\TD{} \WTMM{} method.

Previous application of the \UD{} \WTMM{}
method to \UD{} surrogate dissipation data 
has already revealed the fact that the cancellation exponent
might be different from zero~\cite{tRou96}.
The \TD{} \WTMM{} results reported here show that this 
surprising result is not some artefact resulting from the analysis of 
\UD{} cuts 
(the dissipation could well be non conserved along these
 cuts), but that the multifractal spatial
 structure of the \TD{} dissipation field
is likely to be well described by a multiplicative cascade process that is
definitely non conservative.
As shown in Fig. \ref{fig3}(d), this conclusion is confirmed by the
 results obtained when averaging over 18 snapshots of L\'ev\^eque's DNS;
 the cancellation exponent is found even more negative
 $\tau_\epsilon^{WT}(q=1)+3 = -0.26$, as an indication (as regards to
 the smaller  $R_\lambda$ value) that this exponent
 might decrease to zero in the limit of infinite Reynolds number. To conclude, let us point out that 
the $f_{\epsilon}^{WT}(\alpha)$ data seem to be
even better fitted by a parabola,
as predicted for
 non-conservative log-normal cascade processes. 

We are very grateful to M. Meneguzzi and E. L\'ev\^eque for allowing us
to have access to their DNS data and to the 
CNRS under GDR turbulence.

\end{document}